\documentclass[twocolumn,showpacs,prl]{revtex4}

\usepackage{graphicx}%
\usepackage{dcolumn}
\usepackage{amsmath}
\usepackage{latexsym}

\begin {document}

\title
{
Particle-hole symmetry in a sandpile model
}
\author
{
R. Karmakar and S. S. Manna
}
\affiliation
{
Satyendra Nath Bose National Centre for Basic Sciences
    Block-JD, Sector-III, Salt Lake, Kolkata-700098, India
}

\begin{abstract}
      In a sandpile model addition of a hole is defined as the removal of
   a grain from the sandpile. We show that hole avalanches can be
   defined very similar to particle avalanches. A combined particle-hole
   sandpile model is then defined where particle avalanches are 
   created with probability $p$ and hole avalanches are created with 
   the probability $1-p$. It is observed that the system is critical
   with respect to either particle or hole avalanches for all values
   of $p$ except at the symmetric point of $p_c=1/2$. However at $p_c$
   the fluctuating mass density is having non-trivial correlations
   characterized by $1/f$ type of power spectrum.
\end{abstract}
\maketitle

    The dynamics of a large number of physical processes are characterized
    by bursts of activity in the form of avalanches. For
    example, the mechanical energy release during earthquakes \cite {Earthquake}, 
    river networks  \cite {River}, forest fires  \cite {Forest},
    land slides on mountains
    or sand avalanches on sandpiles etc. Bak, Tang and Wiesenfeld in
    1987 proposed that these systems may actually be exhibiting signatures 
    of a critical stationary state  \cite {Bak}. More precisely they suggested that
    as an indication of the critical state, long ranged spatio-temporal 
    correlations may emerge in some systems governed by a self-organizing
    dynamics, in absence of a fine tuning parameter. This is in essence
    the basic idea of Self-organized Criticality (SOC). 
    Sandpile models are the prototype models of SOC
    \cite {Bak,Dhar,Kad,Grass,Maya,Prie}.

    In the sandpile model an integer variable $n$ representing the number 
    of particles (sand grains) in a sand column is associated with every site
    of a square lattice. The system is driven by adding a particle at a randomly 
    selected site $i$: $n_i \to n_i + 1$.
    A threshold number $n^p_c$ for the stability of a sand column is pre-assigned.
    If at any site the particle 
    number $n_i > n^p_c$ the column topples and this site looses 4 particles 
    and all four neighboring sites $j$ get one particle each \cite {Bak}. 
\begin{equation}
n_i \to n_i - 4 {\hspace*{1.0 cm}}{\rm and} {\hspace*{1.0 cm}} n_j \to n_j+1 
\end{equation}
    As a result some of these neighboring sites may also topple which creates an 
    avalanche of sand column topplings. The extent of such cascading activity 
    measures the size of the avalanche.

        Sand particles drop out of the system through the boundary of the lattice so that in the steady state the fluxes of
    in-flowing and out-flowing particle currents balance. In a stable state number of particles at all
    sites are less than $n^p_c$. Addition of a particle takes the system from one stable state
    to another stable state. Dhar had shown that under this sandpile dynamics, a system evolves 
    to a stationary state where all stable states are restricted to a subset of all possible
    stable states. These states are called recurrent states and they are characterized by the
    absence of forbidden sub-configurations (FSCs) \cite {Dhar1}. All recurrent states occur with
    uniform probabilities in the stationary state. A stable state which is not a recurrent state
    is called a transient state and never appears in the stationary state.

      The main question we would like to ask in this paper is, for an arbitrary sandpile model
   to attain the BTW critical behaviour is it absolutely necessary that the stationary states
   should only be the recurrent states of the BTW model? Can it happen that the neighbouring transient states which
   are very close to the recurrent states of the BTW model are also acceptable in the stationary
   states to achieve the BTW critical behaviour? In the following we introduce the
   concept of holes and on addition of holes to the system the stationary states of the
   resulting sandpile model cannot be anymore strictly restricted to the recurrent states of the BTW model
   since the FSCs can very well be present in the stationary states of this model. In fact
   the recurrent and stationary sets coincide in this model. However the distribution of
   weights of these states may be quite non-trivial and this question remains open. Our numerical results
   show that even in such a case the critical behaviour is very similar to that of the BTW model.

\begin{figure}[top]
\begin{center}
\includegraphics[width=6.5cm]{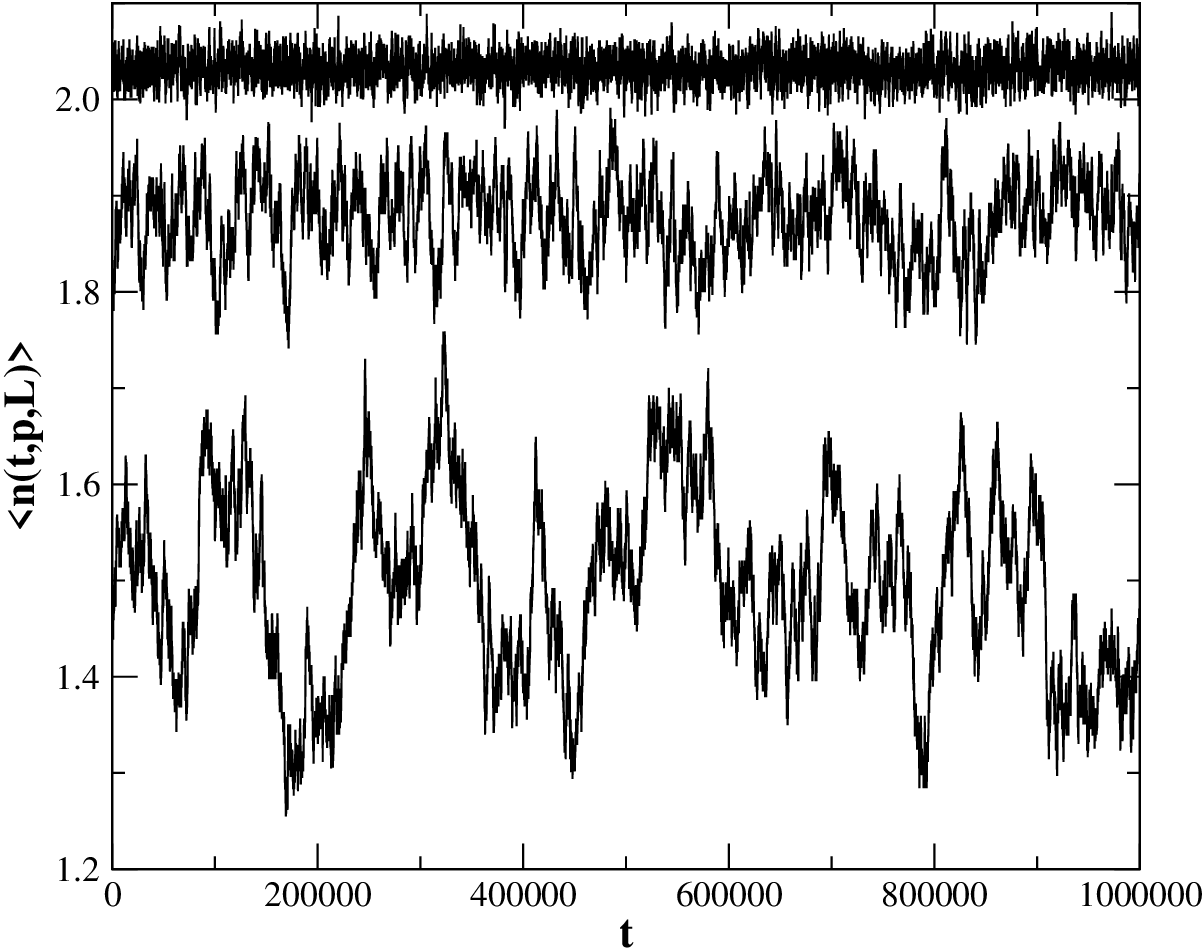}
\end{center}
\caption{
The fluctuation of the mean number of particles per site with time in a system of size $L=64$ for the probabilities
$p=0.60$ (top) 0.52 (middle) and 0.50 (bottom). Both the width of fluctuation and 
the correlation increases as $p$ approaches $p_c=1/2$.
}
\end{figure}

       A `hole' may be defined as the absence of a particle. Therefore adding a
    hole to a lattice site implies taking out one particle from that site: $n_i \to n_i - 1$.
    Repeated addition of holes at randomly selected sites may reduce the number of particles at a
    site less than another pre-assigned threshold $n^h_c$. Therefore
    if $n_i$ at a site is less than $n^h_c$, the site losses four holes i.e., four particles are added
    to this site and each neighboring sites gets one hole (looses one particle):
\begin{equation}
n_i \to n_i + 4 {\hspace*{1.0 cm}}{\rm and} {\hspace*{1.0 cm}} n_j \to n_j-1
\end{equation}
    We call this event as a `reverse toppling'.
    Consequently at some of the neighboring sites particle numbers may also go below the $n^h_c$
    which again reverse topple and thus an avalanche of reverse topplings take place in the system.
    Addition of a particle creates a particle avalanche where as the addition of a hole creates
    a hole avalanche. We assign $n^p_c=3$ and $n^h_c=0$.

    Inverse avalanches were introduced before to get back the recursive configuration
    corresponding to the particle deletion operator \cite {Dhar-Manna}. 

\begin{figure}[top]
\begin{center}
\includegraphics[width=8.0cm]{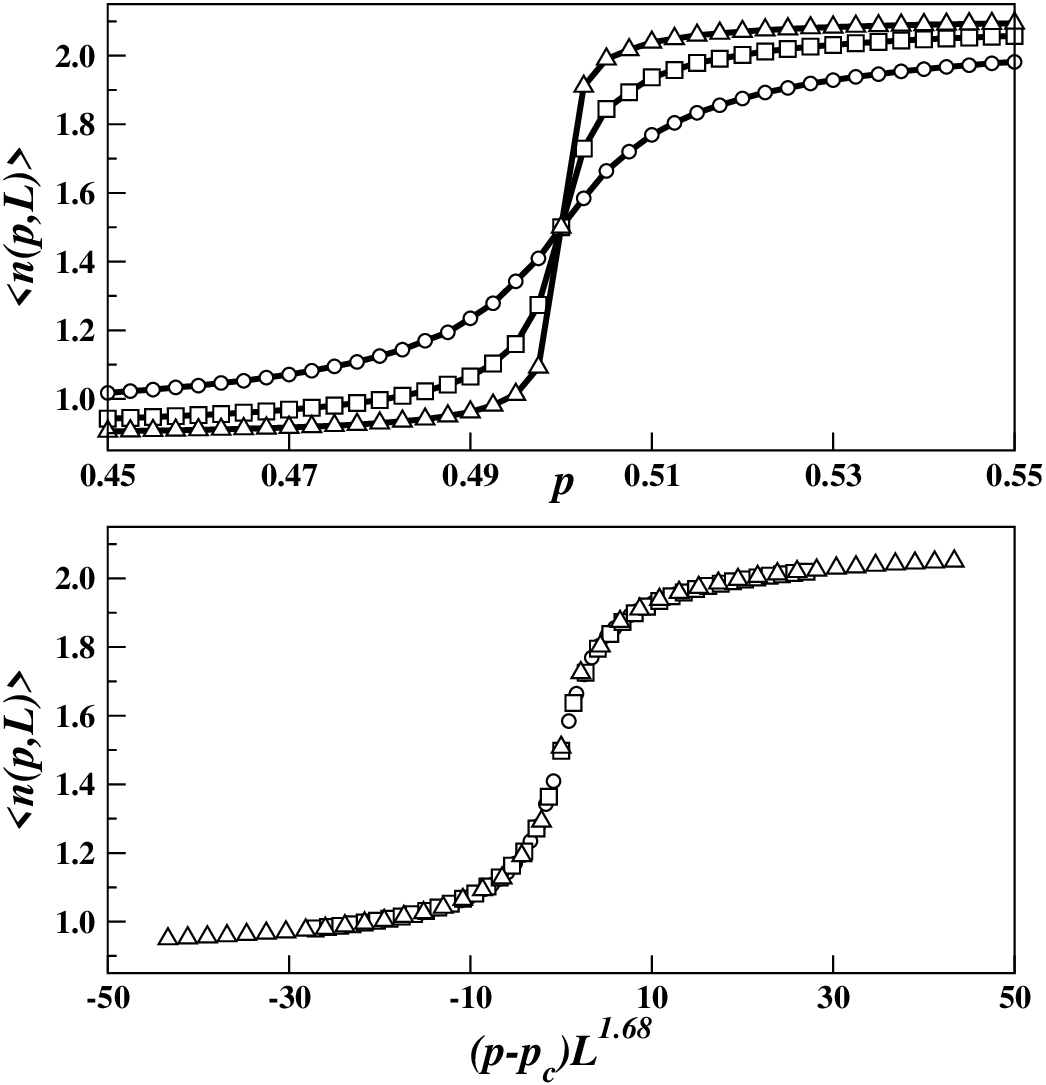}
\end{center}
\caption{
The time averaged number of particles per site $\langle n(p,L) \rangle $ in the stable stationary states
as a function of the probability of
adding a particle $p$. This variation is symmetric about the mid point $p_c=1/2$ and 
$\langle n(p_c,L) \rangle = 3/2$. The data is for $L=32$ (circle), $L=64$ (square) and
for $L=128$ (triangle).
}
\end{figure}

       During the particle avalanches particle current flows into the system by addition of particles
    in the bulk of the system and then they flow out of the system through the boundary. On the contrary
    in hole avalanches particle current flows into the system through the boundary and 
    flows out of the system through the bulk of the system.

       In this paper we study a combination of particle and hole avalanches.
    We probabilistically add either a particle with a probability $p$ or add a hole
    with a probability $1-p$. Therefore when $p=1$, the situation is identical to the ordinary BTW
    model of sand avalanches when no hole is added. On the other hand for $p=0$ only holes are 
    added to the system and no particle. Therefore for $p > 1/2$ more particles are added to the
    system than the number of holes and therefore the net particle current is $2p-1$ and it flows from the bulk of the
    system to the boundary. However for $p < 1/2$ holes are dropped more than particles and the net particle current
    flows in the opposite direction. At $p=1/2$ there is no net current in the
    system. We consider $p=p_c=1/2$ is a critical probability and study the behavior of this
    system around this critical probability. The time $t$ is measured by the number of particles and holes dropped in the system.

      If $n_i(t,p,L)$ is the generalised notation for the number of particles at site $i$, then the
   total number of particles in the system is: $n(t,p,L)=\Sigma^{L^2}_{i=1}n_i(t,p,L)$.
   The mean number of particles per site is then $\langle n(t,p,L) \rangle = n(t,p,L)/L^2$.
   In the stationary state $\langle n(t,p,L) \rangle $ fluctuates 
   rapidly around its time averaged value $\langle n(p,L) \rangle$. 
   These fluctuations are shown in Fig. 1 for the system size $L=64$ and for $p$ = 0.60, 0.52 and
   for 0.50. It is observed from this figure that both the width as well as the correlation
   of fluctuation increases as $p$ approaches $p_c$ from either side of it.

\begin{figure}[top]
\begin{center}
\includegraphics[width=7.5cm]{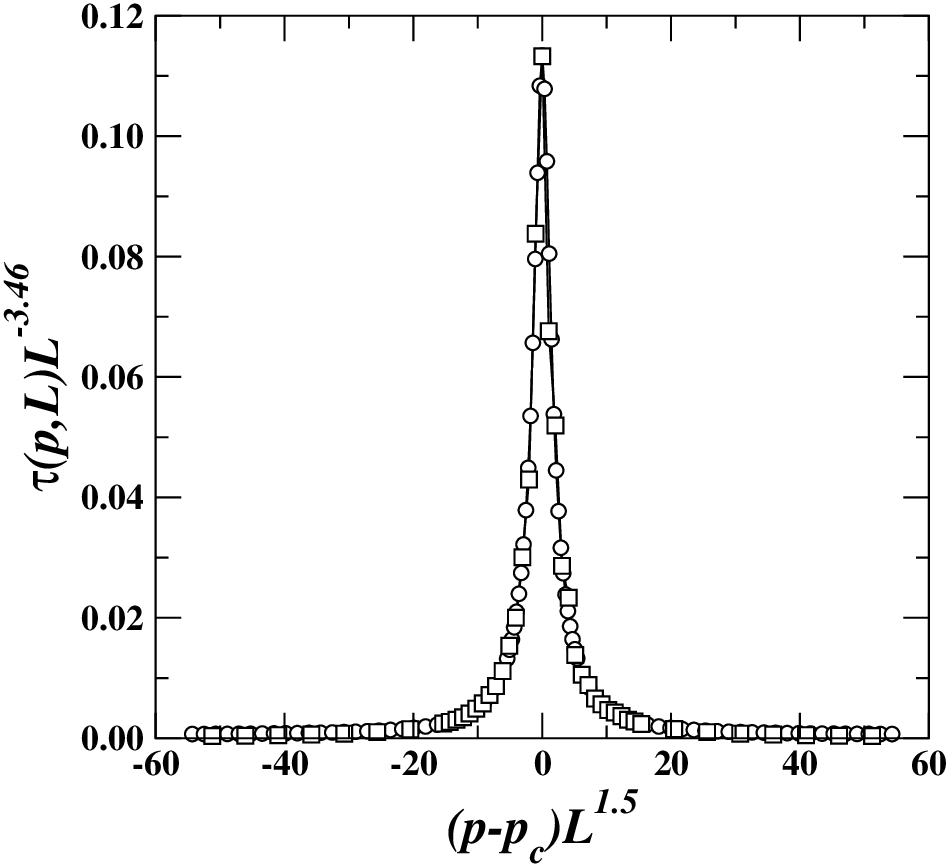}
\end{center}
\caption{
Variation of the correlation time $\tau(p,L)$ of fluctuation of the particle density $\langle n(p,L) \rangle$
with the probability $p$ for system size $L$. The data is shown for two system sizes:
$L=$ 32 and 64.
}
\end{figure}
       The time averaged number of particles per site $\langle n(p,L) \rangle $ is a function of $p$ and the
    system size $L$. At $p=1$
    it is equal to the average number of particles per site in the ordinary BTW model which is
    $n_1 = 2.125$ in the asymptotic limit of large system sizes \cite {Prie,Manna}. 
    As $p$ decreases $\langle n(p,L) \rangle $
    slowly decreases but near $p_c=1/2$ it decreases very fast to a value of $\langle n(p,L) \rangle$ =3/2.
    When $p$ decreases from $1/2$ even further, $\langle n(p,L) \rangle $ decreases fast but eventually
    saturates to a value of $n_0 = 3 - n_1 = 0.875$. In Fig. 2(a) we show this variation. To see if the
    steep rise of $\langle n(p,L) \rangle$ around $p_c$ is associated with some critical exponent,
    we make a scaling plot of $\langle n(p,L) \rangle$ with $p-p_c$ for a number
    of different system sizes $L$ in Fig. 2(b). The data collapse shows:
\begin {equation}
\langle n(p,L) \rangle \sim {\cal G}((p-p_c)L^{1.68}).
\end {equation}

\begin{figure}[top]
\begin{center}
\includegraphics[width=7.5cm]{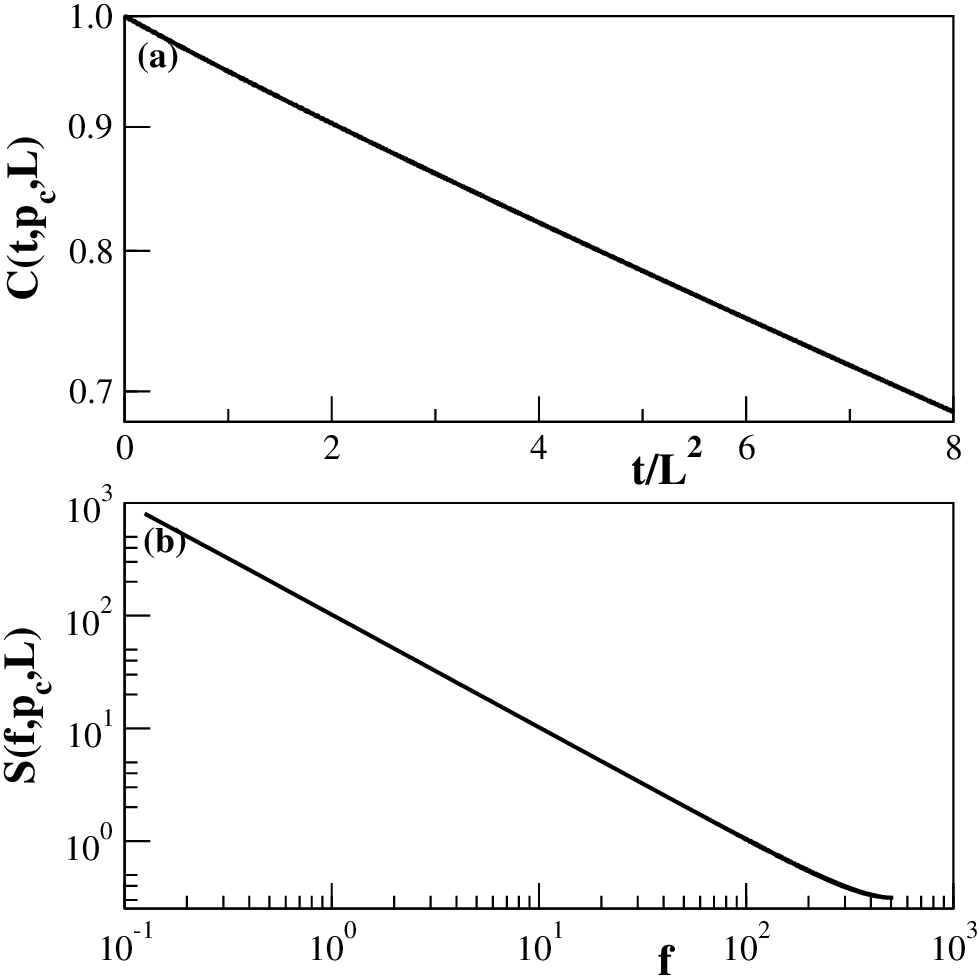}
\end{center}
\caption{(a) The autocorrelation $C(t,p_c,L)$ of the time series of the fluctuating mean number of particles per site 
$\langle n(t,p_c,L) \rangle$
for the system size $L=64$ at $p_c=1/2$. $C(t,p_c,L)$ is plotted with
the scaled time axis $t/L^2$ on a semi-log scale. In (b) the power spectrum $S(f,p_c,L)$ is plotted with
the frequency $f$ on a double logarithmic scale showing a power law decay of the spectrum with
the spectra exponent being nearly equal to one.
}
\end{figure}

      The width of fluctuation is calculated as: $w(p,L) = \langle n^2(p,L) \rangle - \langle n(p,L) \rangle^2$.
   For a given $L$ the width is maximum at $p=p_c$ and then monotonically decreases as $|p-p_c|$ increases.
   On the other hand for a given $p$ the width also decreases with increasing $L$. It is observed
   from numerical estimation that at $p=p_c$, $w(p_c,L)$ 
   decreases with system size as $w(p_c,L) = w_o+w_1L^{-1/2}$ where $w_o = 0.076$
   and $w_1=0.527$ are estimated. Beyond $p_c$ the width decreases as: $w(p,L) = |p-p_c|^{-\nu}$ with 
   $\nu \approx 0.82$ is estimated.

   The time-displaced autocorrelation of the fluctuating mass per site is defined as:
\begin {equation}
C(t,p,L) = \frac {\langle n(t_o+t,p,L)n(t_o,p,L) \rangle - \langle n(p,L) \rangle^2}
{\langle n^2(p,L) \rangle - \langle n(p,L) \rangle^2}
\end {equation}
    This autocorrelation is observed to decay exponentially as: $C(t,p,L) \sim exp(-t/\tau(p,L))$
    where $\tau(p,L)$ is the correlation time. On a semi-log plot of $C(t,p,L)$ vs. $t$ the 
    slope of the plot gives the value of the correlation time $\tau(p,L)$ which is measured
    for different probabilities $p$ and for different $L$ values. For a given system size the correlation time is
    maximum at $p_c$ and then decreases monotonically with increasing $|p-p_c|$. Also 
    $\tau(p,L)$ increases with $L$ at a given $p$. A scaling plot of the
    data collapses very nicely as (Fig. 3):
\begin {equation}
\tau(p,L)L^{-3.46} \sim {\cal F}((p-p_c)L^{1.5})
\end {equation}
    At $p_c$, $\tau(p_c,L)$ increases as $L^{\mu}$ where $\mu$ is estimated
    to be $3.45 \pm 0.10 $.

    Fourier transform of the  autocorrelation function $C(t,p,L)$ is known as the spectral density or
    power spectrum $S(f,p,L)$ defined as
\begin {equation}
 S(f,p,L) = \int^{\infty}_{-\infty} e^{-ift}C(t,p,L)dt
\end {equation}
    In Fig. 4(a) we show the plot of the  autocorrelation function  $C(t,p_c,L)$
    with scaled time $t/L^2$ for a system size $L=64$ and exactly at $p_c$. A straight line plot on a semi-log scale
    implies an exponential decay of the correlation function. In Fig. 4(b) we show the
    Fourier transform of the autocorrelation function plotted in Fig. 4(a) generated
    by the plotting routine `{\it xmgrace}'. On a double logarithmic scale
    the power spectrum $S(f,p_c,L)$ vs. the frequency $f$ plot gives a very good straight
    line for the intermediate range of frequencies implying a power law decay of the
    spectral density: $S(f,p_c,L) \sim f^{-\beta}$. We estimate $\beta \approx 1$ showing the existence of $1/f$
    type of noise in the power spectrum.
 
       The avalanche size distributions for both particle as well as hole avalanches
    are measured. It is observed that in the range of $p > p_c$ the particle avalanche sizes are of widely varying magnitudes
    and of all length scales where as the hole avalanche sizes are very small and of the
    order of unity. Opposite is the situation for the range $p < p_c$. 
    At $p_c$ however both the particle as well as hole avalanche size distributions are similar
    and they are found to follow a stretched exponential distribution like:
\begin {equation}
P(s) \sim \exp(-a s^{\gamma})
\end {equation}
    where $\gamma$ is estimated to be around 0.4. Away from this critical point $p_c$, the particle
    avalanches have power law distribution for $p > p_c$ and hole avalanches follow power law distributions for
    $p < p_c$. Particle avalanche size distributions are calculated at $p=0.51$ and for
    system sizes $L=256, 512, 1024$ and 2048. These distributions are
    very similar to the avalanche distributions in BTW model. For small avalanche sizes they do follow a
    power law distribution $P(s) \sim s^{-\tau}$ where the exponent $\tau$ slowly varies with the
    system size and gradually increases towards 1.2. The large avalanches have multi-fractal
    distribution and simple scaling does not work for the full distribution \cite{Stella1,Stella2}.
    Also the average avalanche size, area and life times have system size dependances very similar to those
    in the BTW sandpile:
    $\langle s(L) \rangle \sim L^2$, $\langle a(L) \rangle \sim L^{1.72}$ and $\langle t(L) \rangle \sim L$.

      To summarize, we have studied a new sandpile model where both particle as well as hole
   avalanches are created. Their relative stengths are tuned by a parameter $p$ varying between 0 and 1 
   which is the probability for adding a particle and consequently $1-p$ being the hole addition probability.
   Specifically at $p=1$ the system is identical to the ordinary
   BTW model for only particle avalanches. Similarly at $p=0$ there is only hole avalanches and
   their distribution are very similar to the avalanche size distribution for the BTW model.
   In the range $1/2 < p < 1$ there are particle as well as hole avalanches, but the net current
   is due to the particles which flows into the bulk of the system. Critical behavior of the
   particle avalanches are observed to have multi-scaling behavior and is very similar to those of the BTW model.
   Opposite situation happens in the range $0 < p < 1/2$ where net current is due to holes
   which flows into the bulk of the system. The hole avalanche sizes also have multi-scaling distributions
   very similar to the BTW model.

      We thankfully acknowledge S. M. Bhattacharjee for useful discussions.

\end{document}